\documentclass[twocolumn,aps,prb,floatfix,showpacs]{revtex4}
\usepackage{epsfig}% Include figure files
\usepackage{dcolumn}% Align table columns on decimal point
\usepackage{bm}% bold math
\begin{document}
\title{Upper critical fields and thermally-activated transport  of Nd(O$_{0.7}$F$_{0.3}$)FeAs  single crystal.}

\author{J. Jaroszynski,  F. Hunte, L. Balicas, Youn-jung Jo,
I. Rai\v{c}evi\'{c},  A. Gurevich, D. C. Larbalestier}

\affiliation{National High Magnetic Field Laboratory, Florida
State University, Tallahassee, Florida 32310, USA}

\author{F. F. Balakirev}

\affiliation{National High Magnetic Field Laboratory, Los Alamos
National Laboratory, Los Alamos, New Mexico 87545, USA}

\author{L. Fang, P. Cheng, Y. Jia and H. H. Wen}

\affiliation{Institute of Physics, Chinese Academy of Sciences,
Beijing 100190, People's Republic of China}

\date{\today}

%%%%%%%%%%%%%%%%%%%%%%%%%%%%%%%%%%%%%%%%%%%%%%%%%%%%%%%%%%%%%
\begin{abstract}

 We present detailed measurements of the longitudinal
 resistivity $\rho_{xx}(T,H)$, and the upper critical field $H_{c2}$ of
 Nd(O$_{0.7}$F$_{0.3})$FeAs single crystals in strong DC and
 pulsed magnetic fields up to 45~T and 60~T, respectively.
 We found that the  field scale of $H_{c2}$ is
 comparable to $H_{c2}\sim100$~T  of high $T_c$ cuprates.
  $H_{c2}(T)$ parallel to the
 c-axis exhibits a pronounced upward curvature similar to what was
 extracted from earlier measurements on polycrystalline La(O,F)FeAs,
 Nd(O,F)FeAs and  Sm(O,F)FeAs samples.
 Thus this behavior of  $H_{c2}^{\perp}(T)$  is indeed an intrinsic feature of
 oxypnictides, rather than manifestation of vortex lattice melting or granularity.
 The orientational dependence of $H_{c2}(\theta)$ as a function of
 the angle $\theta$ between $\bm{H}$ and the c-axis shows deviations
 from the  one-band Ginzburg-Landau scaling.
 The mass anisotropy parameter $\gamma(T) =(m_c/m_{ab})^{1/2}
 =H_{c2}^\|/H_{c2}^\perp$ obtained from these measurements decreases
 as temperature decreases, from $\gamma\simeq 9.2$ at $44$~K to
 $\gamma\simeq 5$ at $34$~K, where $||$ and $\perp$ correspond to $\bm{H}$ parallel
 and perpendicular to the ab planes, respectively.
 Spin dependent  magnetoresistance and nonlinearities in the Hall
 coefficient suggest contribution to the conductivity
 from electron-electron  interactions modified by disorder reminiscent
  that of  diluted magnetic semiconductors.
  The Ohmic resistivity $\rho_{xx}(T,H)$
 measured below $T_c$ but above the irreversibility field
  exhibits a clear Arrhenius thermally
 activated behavior $\rho=\rho_0\exp[-E_a(T,H)/T]$ over $4-5$
  decades of $\rho_{xx}$.
The activation energy $E_a(T,H)$ has very different field
dependencies for $\bm{H}||\bm{ab}$ and $\bm{H}\perp \bm{ab}$,
varying from $4\times 10^3$~K at $H=0.2$~T  to $\sim 200$~K at
$H=35$~T. We discuss to what extent different pairing scenarios
suggested in the literature can manifest themselves in the
observed behavior of $H_{c2}$, using the two-band model of
superconductivity in oxypnictides. The results indicate  the
importance of paramagnetic effects on $H_{c2}(T)$ in oxypnictides,
which may significantly reduce  $H_{c2}(0)$ as compared to
$H_{c2}(0)\sim200-300$~T based on extrapolations of  $H_{c2}(T)$
near $T_c$ down to low temperatures.

\end{abstract}

%%%%%%%%%%%%%%%%%%%%%%%%%%%%%%%%%%%%%%%%%%%%%%%%%%%%%%%%%%%%%

\maketitle

The recently discovered layered superconducting oxypnictides with
high transition temperatures $T_c$ \cite{hosono08} are based on
alternating structures of FeAs layers and REO layers of rare earth
(RE) oxides. Similar to the high-temperature superconducting
cuprates, superconductivity in oxypnictides emerges upon doping of
a parent antiferromagnetic state. As the REO planes are doped, the
ionically bonded REO donates an electron to the covalently bonded
FeAs plane \cite{Nature}, suppressing the global
antiferromagnetism and resulting in superconductivity.

Previous low field \cite{lowfield1,lowfield2,lowfield3,lowfield4}
and high field transport studies \cite{Hunte,jjprb08} on different
polycrystalline oxypnictides have shown, that their
magneto-transport behavior is intermediate between conventional
low-$T_c$ superconductors and the high-temperature cuprates.
However, while La(O,F)FeAs behaves as an intermediate-$T_{c}$
superconductor similar to MgB$_{2}$ in which thermal fluctuations
of vortices do not significantly affect the $H$--$T$ diagram to
the extent that they do in the layered cuprates, the higher $T_c$
oxypnictides like SmFeAsO$_{0.85}$ and Nd(O,F)FeAs exhibit larger
mass anisotropies, enhanced thermal fluctuations, and a Ginzburg
parameter comparable to that of YBCO. Thus, the question arises,
whether this distinct behavior stems from different critical
temperatures, the presence of magnetic ions \cite{mag1,mag2,mag3},
or is due to the electromagnetic granularity of these
polycrystalline samples \cite{yamamotoLa,yamamotoSm,proz}. For
example, is the pronounced upward curvature of $H_{c2}(T)$ along
the c-axis extracted from measurements of the resistivity onsets
of polycrystals \cite{Hunte}  indeed an intrinsic feature of
$H_{c2}^\perp(T)$ in oxypnicties, or does it reflect an extrinsic
effect of magnetic granularity, the onset of the irreversible
critical state behavior and/or melting of the vortex lattice?
Moreover, given that the nature of superconductivity in the
oxypnictides is still poorly understood, several different
scenarios have been proposed. In particular, many theoretical
models are based on the multiband electronic structure of the
oxypnictides and the fact that the superconducting state emerges
upon doping of the parent antiferromagnetic semimetal
\cite{mazin,theor1,theor2,theor3,theor5}. The multiband effects
would manifest themselves in the distinct temperature and
orientational dependencies of $H_{c2}(T)$, as has been well
documented in detailed studies of MgB$_2$
\cite{h1,h2,h3,h4,h5,h6,h7,h8,h9,h10,h11,h12,ag}. To address these
issues, we made detailed measurements of the magnetotransport
properties of a Nd(O$_{0.7}$F$_{0.3}$)FeAs single crystal in very
high magnetic fields.

\section{Samples}

The crystal was made by the flux method using NaCl as the flux.
First the starting  Nd  (purity 99.95~\%) and As  (purity
99.99~\%) were mixed in 1:1 ratio, ground and pressed into a
pellet shape. Then it was sealed in an evacuated quartz tube and
reacted at 800 $^{\circ}$C for 10 hours. The resultant pellet was
ground together with a stoichiometric powder mixture of  NdF$_3$
(purity 99.95~\%), Fe$_2$O$_3$  (purity 99.9~\%) and Fe  (purity
99.9\%) resulting in Nd(O$_{0.7}$F$_{0.3}$)FeAs compound. Again it
was pressed into a pellet and put together with NaCl powder of
mass ratio (NaCl : Nd(O,F)FeAs = 10:1) and sealed in an evacuated
quartz tube and reacted at about 1050 $^{\circ}$C for 10 days.
Then it was cooled down at a rate of 3 $^{\circ}$C/hour to 850
$^{\circ}$C, followed by a quick cooling to room temperature by
shutting off the power of the furnace. The resulting product
contains mainly plate-like small crystals with lateral dimensions
of 5--50~$\mu$m and $\sim 1$~$\mu$m thickness.

Previous extensive transport \cite{Wen_Tc,Wen_Hall,Wen_anizo} and
heat capacity \cite{Wen_heat} studies on
Nd(O$_{0.82}$F$_{0.18}$)FeAs single crystal revealed the onset of
superconductivity at $T_c\approx 47$~K, the upper critical fields
$H_{c2}^{\|}(0) = 304$~T and $H_{c2}^{\perp}(0)=62-70$~T
extrapolated from  the Werthamer-Helfand-Hohenberg (WHH)
formula\cite{WHH}. These samples also have a relatively small mass
anisotropy parameter $\gamma = (m_c/m_a)^{1/2}\lesssim5$ where
$m_a$ and $m_c$ are the effective masses along the ab plane and
the c-axis, respectively. However, the above studies were
performed in magnetic fields not exceeding 9~T, where most of the
multi-band effects would not manifest themselves in the
$H_{c2}(T)$ curves.

\section{Resistive transitions and upper critical fields}

\begin{figure}
\centering
\includegraphics[width=7.5cm]{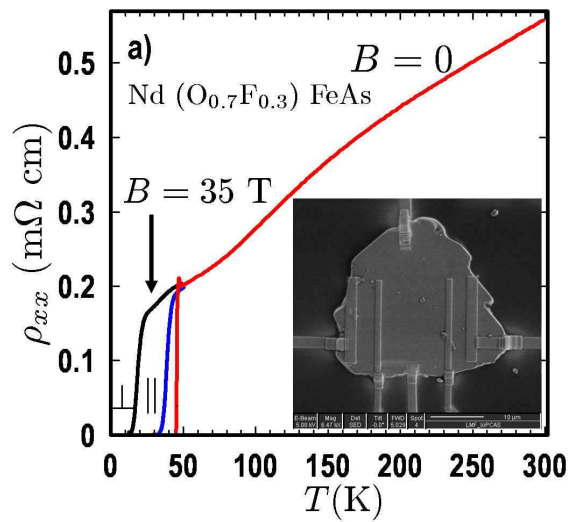}

\caption{Longitudinal resistivity $\rho_{xx}(T)$ at $H=0$ and
$H=35$~T for $H||ab$ and $H\perp ab$. At $B=0$ superconductivity
onset is observed at $T_c\simeq 46$~K. Inset shows scanning
electron micrograph of the sample.} \label{fig:rho1}
\end{figure}

For our experiment we used three different high-field magnets at
the National High Magnetic Field Laboratory (NHMFL): 35~T DC
resistive and 45~T hybrid magnets at Florida State University and
65~T pulsed field magnet at Los Alamos National Laboratory.  The
inset to Fig.~\ref{fig:rho1}(a) shows the scanning electron
micrograph of the Nd(O,F)FeAs sample with six Pt contacts
fabricated by the focused ion beam (FIB) technique. The distance
between voltage contacts is $\sim 13$~$\mu$m, and the sample
thickness is $\sim 1.4$~$\mu$m. The sample was glued to a
conductive silicon substrate to ensure good thermalization during
measurement, and to prevent electric charge build-up during the
FIB procedure. The sample was placed on a rotating platform, which
allows us to change its orientation with respect to the magnetic
field {\em in situ}. The longitudinal resistivity $\rho_{xx}$ and
the Hall coefficient $R_H$ in high magnetic fields were measured
using a lock-in technique with AC excitation  current below 0.3~mA
at frequency 22 Hz.  Pulsed field measurements were conducted
using in-house synchronous digital lock-in technique at frequency
of 16~kHz and similar levels of current.

\begin{figure}
\centering
\includegraphics[width=7.5cm]{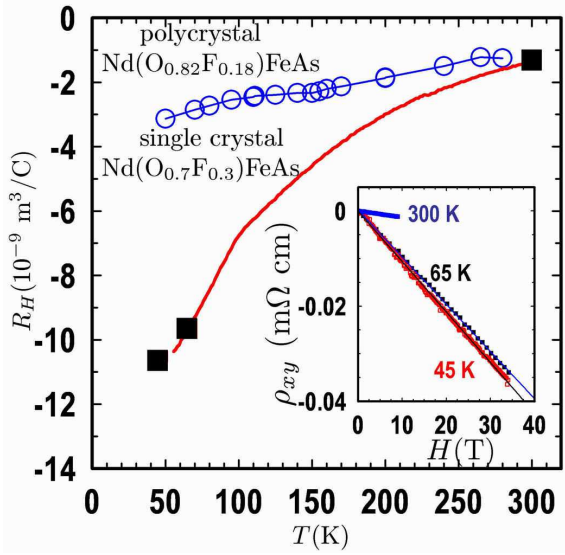}

\caption{Hall coefficient $R_H$ as a function of $T$ measured in
superconducting magnet at $\pm 9$~T (solid line), and determined
from linear fits to $\rho_{xy}(H)$ for $-35<H<35$~T or $-9<H<9$~T
(solid squares), as shown in the inset. $R_H$ for polycrystalline
Nd(O$_{0.94}$F$_{0.06}$)FeA sample is also shown. }
\label{fig:rho2}
\end{figure}

Figure~\ref{fig:rho1} shows $\rho_{xx}(T)$ as a function of
temperature $T$ at $H=0$  and $35$~T for different sample
orientations. The magnetic field perpendicular to the ab plane
shifts the superconducting transitions to lower $T$, reducing
$T_c$ (from 46~K to 21~K at $H=35$~T) without significant change
of the shape of $\rho_{xx}(T)$. This behavior is reminiscent of
magnetotransport in low-$T$ superconductors. In contrast, fields
applied parallel to the ab plane change the shape of the resistive
transitions more than they do for $H||c$, but less than what was
observed on polycrystalline Nd(O,F)FeAs \cite{jjprb08}.
Figure~\ref{fig:rho1} shows that the $\rho_{xx}(T)$ dependence is
sublinear at higher $T$ but superlinear below $T\lesssim100$~K.
Over the same temperature range, the Hall resistivity
$\rho_{xy}(T)$ shown in Fig.~\ref{fig:rho2} has a relatively
strong temperature dependence for such an optimally doped
material. This dependence is much stronger than  was observed on a
polycrystalline sample with much lower doping,
 $x=6$~\%  \cite{jjprb08}, with $\rho_{xy}(T)$ exhibiting
  a kink  around $T\approx100$~K.
 Such temperature dependencies of the transport coefficients near
$\sim100$~K may result from a multiband conductivity or a
structural transition. The inset to Fig.~\ref{fig:rho2} shows
$\rho_{xy}(H)$ at selected temperatures. On this field scale,
$\rho_{xy}(H)$ does not exhibit  field nonlinearities, at
$40<T<300$K, similarly to  polycrystalline samples \cite{jjprb08}.
However,  a strong dependence of $R_H(H)$ is observed in this
single crystal at lower temperatures as will be discussed later.

\begin{figure}
\centering
\includegraphics[width=7.5cm]{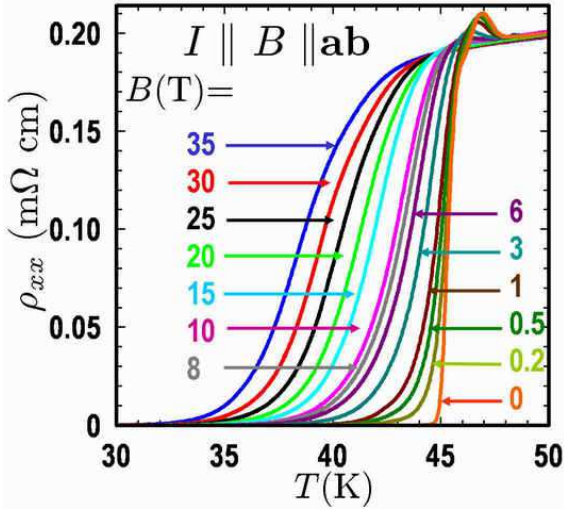}

\caption{$\rho_{xx}$ as a function of temperature at various
magnetic fields for $H\|  ab$.} \label{fig:rho3}
\end{figure}

\begin{figure}
\centering
\includegraphics[width=7.5cm]{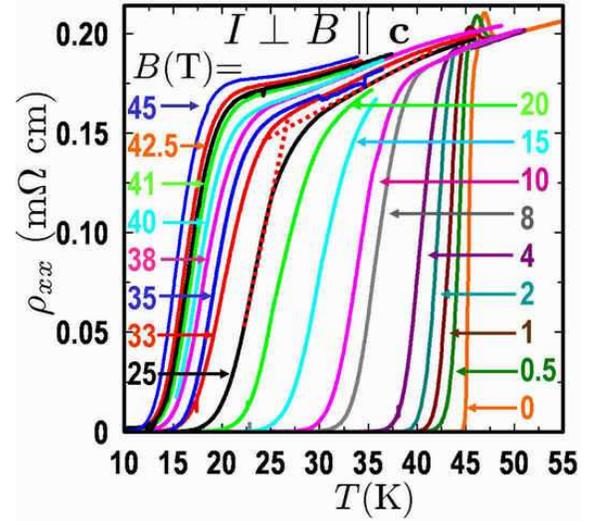}

\caption{$\rho_{xx}$ as a function of temperature at various
magnetic fields for $H\perp ab$. Dotted lines  show how the onset
of the transition was determined for  $H=20$~T. Excitation current
$I\leq0.3$~mA, at frequency $f=22$~Hz was applied in the {\bf ab}
plane.} \label{fig:rho4}
\end{figure}

\begin{figure}
\centering
\includegraphics[width=7.5cm]{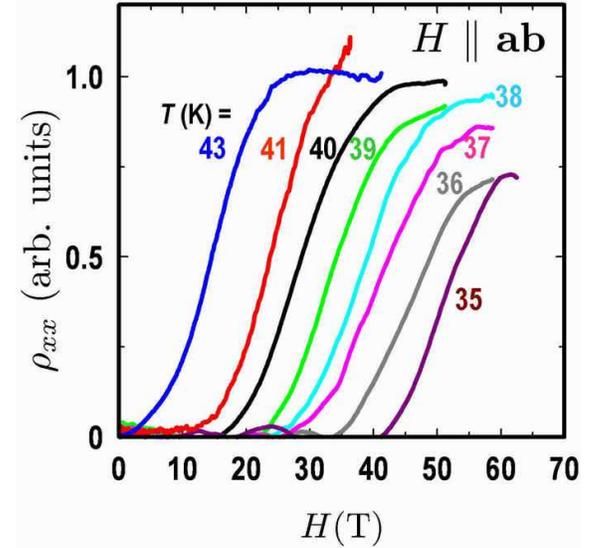}

\caption{ The longitudinal resistivity $\rho_{xx}(H)$ as a
function of  magnetic field in for $H\| ab$ measured in pulsed
magnet up to 60~T.   } \label{fig:mr1}
\end{figure}

\begin{figure}
\centering
\includegraphics[width=7.5cm]{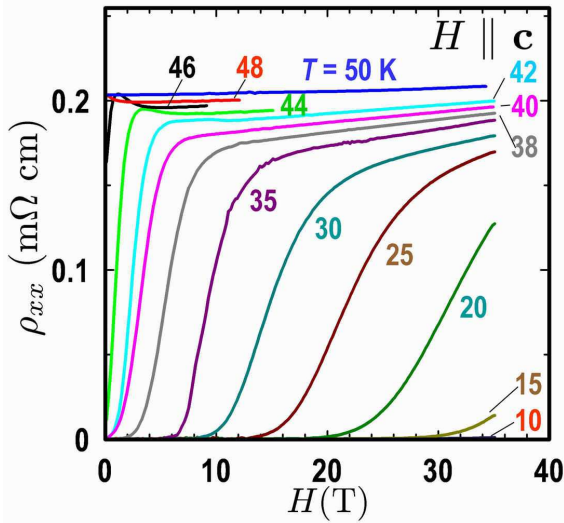}

\caption{ The longitudinal resistivity $\rho_{xx}(H)$ as a
function of  magnetic field in for $H\perp ab$ measured in DC
resistive magnet up to 35~T at various temperatures.  }
\label{fig:mr2}
\end{figure}

\begin{figure}
\centering
\includegraphics[width=7.5cm]{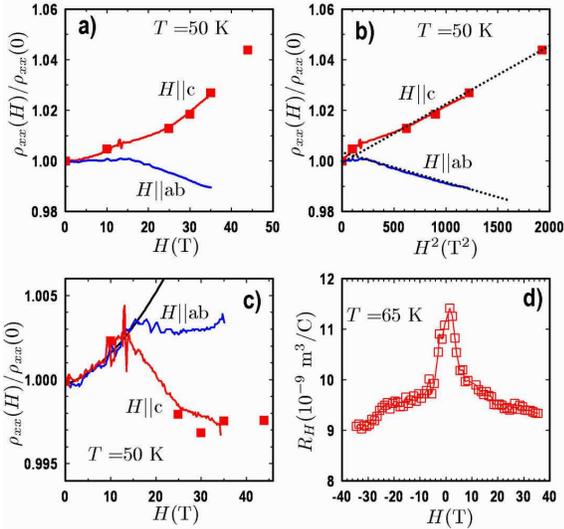}

\caption{Magnetoresistance measured at $T=50$~K for two
different sample configurations. Squares denote data from
temperature sweeps at fixed magnetic field, while solid lines were
measured at fixed $T$. (b) The same data plotted versus $H^2$.
Dotted lines show linear fits taken for $H\gtrsim15$~T. (c) Data
plotted after subtraction of the $H^2$ contribution. (d) Hall
coefficient $R_H(H)=\rho_{xy}(H)/H$ versus magnetic field measured
at $T=65$~K .}  \label{fig:mr3}
\end{figure}

\begin{figure}
\includegraphics[width=8cm]{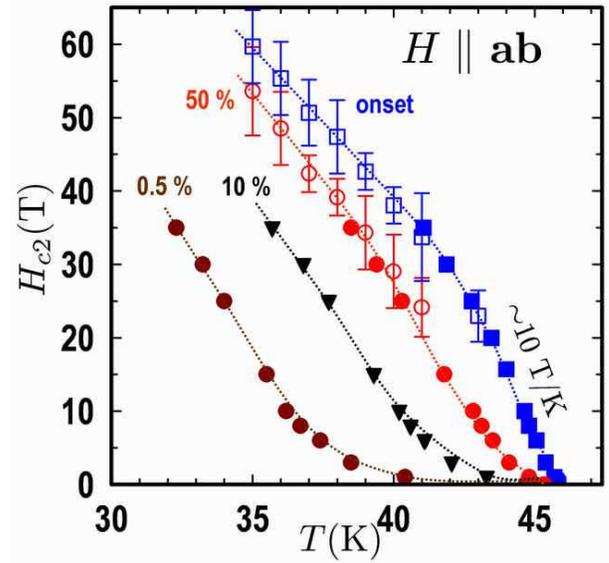}

\caption{Temperature dependence of the upper critical field $H^{||}_{c2}(T)$
along the ab plane obtained from dc (filed symbols) and pulsed
(open symbols) measurements.
The data extracted from the results shown in Fig.~\ref{fig:rho3},
 and Fig.~\ref{fig:mr1}, show the temperatures
at which the resistance reaches 0.5~\%, 10~\%, 50~\% of the normal
state resistance, as extrapolated linearly from the
$\rho_{N}(T,H)$ temperature dependence above $T_c(H)$. The onsets
of superconducting transition were determined as shown in
Fig.~\ref{fig:rho4} and correspond well to the 90~\% transitions.
The dotted lines are guides to the eye. } \label{fig:Bc2ab}
\end{figure}

\begin{figure}
\includegraphics[width=8cm]{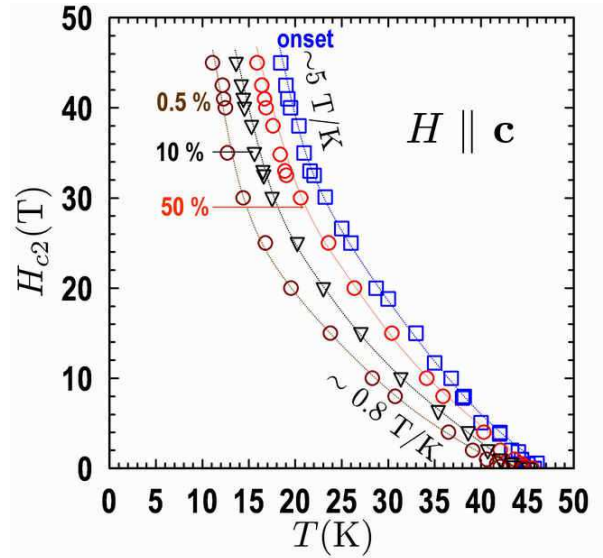}

\caption{Temperature dependence of the upper critical fields $H^{\perp}_{c2}(T)$
along the c-axis obtained from the dc measurements. The data extracted from the results shown in
Fig.~\ref{fig:rho4}, and Fig.~\ref{fig:mr2}, show the temperatures
at which the resistance reaches 0.5~\%, 10~\%, 50~\% of the normal
state resistance, as extrapolated linearly from the
$\rho_{N}(T,H)$ temperature dependence above $T_c(H)$. The onsets
of superconducting transition were determined as shown in
Fig.~\ref{fig:rho4} and correspond well to the 90~\% transitions.
The dashed lines are guides to the eye. } \label{fig:Bc2c}
\end{figure}

Figures~\ref{fig:rho3}--\ref{fig:mr2} show the detailed field and
temperature dependencies of $\rho_{xx}(T,H)$, from which the upper
critical fields parallel and perpendicular to the ab planes were
extracted. Here $H_{c2}$ for both field orientations was measured
at the onset of the superconductivity transition defined by the
intersection point from linear extrapolations of $\rho_{xx}(T,H)$
at $T<T_c$ and $\rho_n(T,H)$ at $T>T_c$, as shown in Fig.
~\ref{fig:rho4}. Here $\rho_n(T,H)$ is the resistivity in the
normal state. We also extracted the mid transition field
$H_{50}(T)$ and the resistivity onset fields $H_{10}(T)$ and
$H_{0.5}(T)$ defined by $\rho_{xx}(T,H)=0.5\rho_n(T,H)$,
$\rho_{xx}(T,H)=0.1\rho_n(T,H)$, and
$\rho_{xx}(T,H)=0.05\rho_n(T,H)$, respectively. The fields
$H_{0.5}(T)$ are close to the irreversibility field $H^*$, which
quantifies the onset of vortex critical state. The resulting
temperature dependencies of all these fields are summarized in
Figs.~\ref{fig:Bc2ab} and \ref{fig:Bc2c}. Notice that the $R(T)$
curves shown in Figs. ~\ref{fig:rho3}--\ref{fig:mr2} exhibit a
small peak just before the transition which disappears at $H>4$T
for ${\bf H}|| c$ and  $H>6$T for ${\bf H}|| ab$. Such resistance
peak anomaly  has been observed in other superconductors and is
usually ascribed to the effect of paramagnetic ions, sample
inhomogeneities, contact arrangements, etc.
\cite{respeak1,respeak2,respeak3,respeak4}. Since the mechanism of
this anomaly is still not completely understood, we did not take
it into account when extracting $H_{c2}$ from the data shown in
Figs. ~\ref{fig:rho3}--\ref{fig:mr2}.

Figure \ref{fig:mr3}(a) shows the magnetoresistance (MR) at
$T=50$~K, just above the onset of superconductivity. The  MR is
positive for $H||$c  and negative for for $H||$ab. The difference
could be attributed to orbital origin of this MR. As shown in
Fig.~\ref{fig:mr3}(b), the magnetoresistance,
$(\rho_{xx}(H)-\rho_{xx}(0))/\rho_{xx}(0)\propto H^2$ for both
field orientations and $H\gtrsim15$~T exhibits a quadratic field
dependence.  However, if this quadratic contributions are removed,
a weak remanent positive MR is observed at low fields
$H\lesssim15$~T, as shown in Fig.~\ref{fig:mr3}(c). This remanent
MR is identical for both field orientations, which in turn
suggests a spin mechanism behind this positive MR component.
Interestingly, such a spin dependent  positive MR is ubiquitous in
diluted magnetic semiconductors (DMS) in the paramagnetic phase
\cite{Sawi86,smor97,Andr05}. It originates from the giant
spin-splitting $\Delta_s$ of the electron states, which
considerably affects quantum corrections to the conductivity
brought about by the effect of disorder modified electron-electron
interactions \cite{Alts85Fuku85Lee85}.  Since  $\Delta_s$ in DMSs
is proportional to the magnetization $M$ of the localized spins,
this positive MR scales with $B$ and $T$ like the Brillouin
function. Importantly, this positive MR is absent in a
ferromagnetic phase, when localized moments are already aligned at
$H=0$. This mechanism can also produce MR in Nd(O,F)FeAs, provided
that at least some of localized magnetic moments $\mu$ (most
likely of Nd$^{3+}$ ions with $\mu=3.6\mu_B$)  are in a
paramagnetic phase. At the same time, as shown in
Fig.~\ref{fig:mr3}(d), the Hall coefficient $R_H$ strongly depends
on $H$ at low field. Such a dependence could also result from
quantum corrections to $\rho_{xy}$ from electron-electron
interactions, but maybe a result of multiband conductivity, as
well.

Based on the results shown in Figs.~\ref{fig:Bc2ab} and
\ref{fig:Bc2c}, the following points can be made:

1. The 100--200~T field scale of $H_{c2}$ in
Nd(O$_{0.7}$F$_{0.3})$FeAs is comparable to that of high-$T_c$
cuprates.

 2. $H_{c2}^{\perp}(T)$ perpendicular to the ab plane
exhibits a pronounced upward curvature similar to what was first
reported for polycrystalline La(O,F)FeAs \cite{Hunte}. This is
very different from the one-band WHH behavior and appears to be
 an intrinsic property of oxypnictides rather than
manifestations of the vortex melting or the onset of pinning at
$H=H^*(T)$.

3. The slopes $H_{c2}^{\prime\|}=|dH_{c2}^\|/dT|$ and
$H_{c2}^{\prime\perp}=|dH_{c2}^\perp/dT|$ for both field
orientations increase significantly just a few Kelvins below
$T_c$.  Thus, low-field $H<9$~T measurements may underestimate the
actual values of $H_{c2}^{\prime\perp}=\phi_0/2\pi\xi_{a0}^2T_c$
and $H_{c2}^{\prime\|} =\phi_0/2\pi\xi_{a0}\xi_{c0}T_c$ at
$T\approx T_c$, from which the coherence lengths
$\xi_a(T)=(1-T/T_c)^{-1/2}\xi_{a0}$ and
$\xi_c(T)=(1-T/T_c)^{-1/2}\xi_{c0}$ in the ab plane and along the
c-axis can be extracted. As it is evident from
Figs.~\ref{fig:Bc2ab} and \ref{fig:Bc2c}, the high-temperature
slopes $H_{c2}^{\prime\perp}$ and $H_{c2}^{\prime\|}$ attain
different values above $\simeq 10$~T, so the use of high magnetic
fields becomes decisive for revealing the true intrinsic
superconducting parameters of Nd(O$_{0.7}F_{0.3})$FeAs single
crystals. This may also explain why low-field measurements
\cite{Wen_Tc,Wen_Hall,Wen_anizo} show lower values of
$dH_{c2}^\|/dT$ and $dH_{c2}^\perp/dT$.

\section{Thermally activated resistivity}

\begin{figure}[t]
 \includegraphics[width=7.5cm]{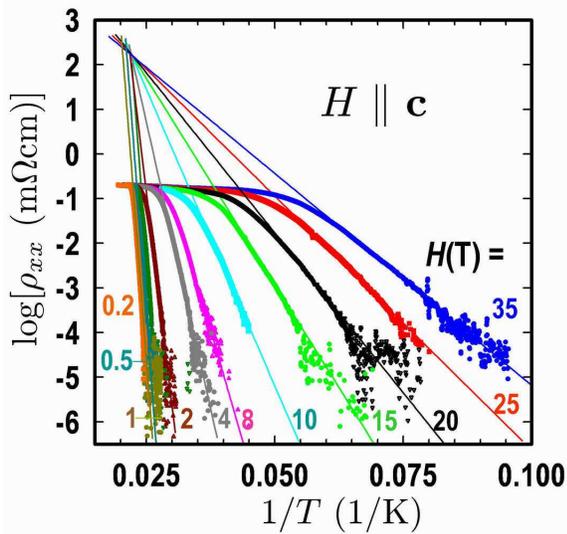}
\centering

\caption{Arrhenius plot for $\rho_{xx}(T)$ at different magnetic
fields perpendicular to the ab plane.}      \label{fig:arr1}
\end{figure}

\begin{figure}[t]
\includegraphics[width=7.5cm]{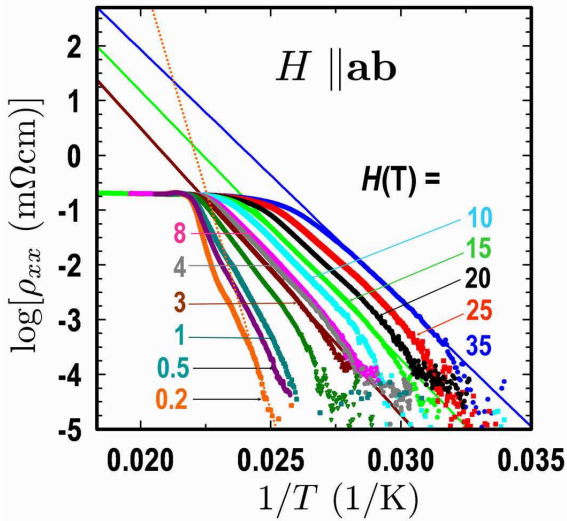}
\centering \caption{Arrhenius plot for $\rho_{xx}(T)$ at different
magnetic fields parallel to the ab plane. }\label{fig:arr2}
\end{figure}

Shown in Figs~\ref{fig:arr1} and  \ref{fig:arr2} are the temperature
dependencies of the ohmic resistivity $\rho_{xx}(T,H)$ measured
above the irreversibility field, $H^*<H<H_{c2}$ for both field
orientations. These data can be described well by the Arrhenius
dependence
    \begin{equation}
    \ln\rho_{xx}=\ln\rho_0-E_a(T,H)/T,
    \label{arr} \\
    \end{equation}
characteristic of thermally-activated flux flow (TAFF) of vortices,
which has been studied extensively on high-$T_c$ cuprates. Here
$E_a(T,B)=-\partial\ln\rho_{xx}/\partial(1/T)$ is the activation
energy determined by hopping of vortex bundles in the pinning
potential. Several features of the data shown in
Figs.~\ref{fig:arr1} and \ref{fig:arr2} should be pointed out.

1. The Arrhenius dependence holds over 4-5 decades of $\rho_{xx}$,
down to the limit of sensitivity of our transport measurements. This
indicates that thermally-activated vortex hopping in
Nd(O$_{0.7}$F$_{0.3})$FeAs is similar to that of high-$T_c$ cuprates
and is much more pronounced than in lower-$T_c$ pnictides like
La(O$_{1-x}$F$_x)$FeAs. The good linear behavior evident from
Figs~\ref{fig:arr1} and  \ref{fig:arr2} also indicates that the
low-temperature dependence of $E_a(T)$ is approximately linear,
$E_a(T)=E_m(1-T/T_m)$, where $T_m$ is a parameter with the
dimensionality of temperature. In this case the term $T/T_m$ can be
eliminated by re-defining $\rho_0\to \rho_0\exp(E_m/T_m)$ in Eq.
(\ref{arr}) without changing any observed characteristics.

2. The higher temperature $\log\rho_{xx}(T)$ data level off at a
field independent value corresponding to the normal state
resistivity $\rho_n$.

3.  The $\log\rho_{xx}(T)$ lines for different $H||c$ extrapolate to
the same temperature, $\approx T_c$, indicating that $T_m=T_c$,
$\rho_0\approx\rho_n\exp(E_mT_c)$ and $E_m$ is practically
independent of $H$.  However, for the parallel field orientation,
this is not the case because of the field dependence of $E_m$, as
shown below.

\begin{figure}[t]
 \includegraphics[width=7.5cm]{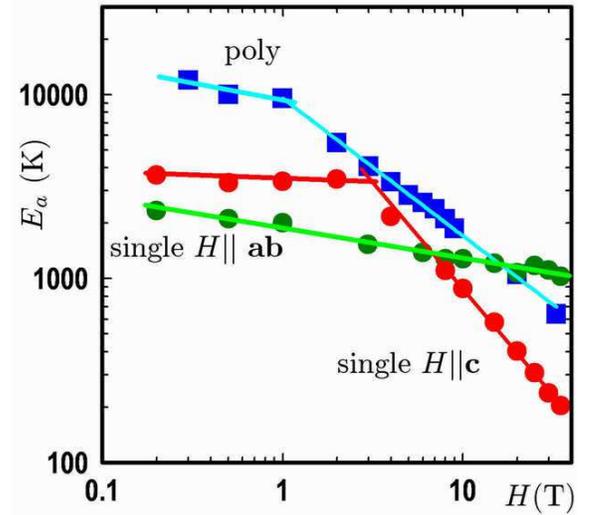}
\centering
\caption{Field dependencies of the activation energy $E_a(H)$ for $H\perp ab$ and
$H\| ab$. The solid squares show the data for Nd(O,F)FeAs polycrystal measured earlier.}
\label{fig:energy}
\end{figure}

From the low-temperature slope of $\log\rho_{xx}$ the field
dependencies of the activation energy
$E_a=-d\mbox{ln}\rho_{xx}/d(1/T)$ can be obtained.  The results
presented in Fig.~\ref{fig:energy} for both field orientations
show that $E_a\simeq (3-4)\times 10^3$~K at low fields. The
activation energy $E_a\simeq 2\times 10^4$~K for a polycrystaline
Nd(O,F)FeAs measured previously \cite{jjprb08} is higher than for
the single crystals studied in this work (similar
thermally-activated resistivity was recently observed on
polycrystalline Nd(O$_{0.82}$F$_{0.18}$)FeAs \cite{lowfield4}).
The field dependencies of $E_a(B)$ for $H||$c and $H\perp$c are
also markedly different. $E_a(H)$ for $H||$c exhibits a rather
conventional field dependence characteristic of
thermally-activated transport: a constant value at low fields
$H<3$ T where single-vortex pinning dominates, followed by a power
law decrease $E_m\propto H^{-1.1}$ characteristic of collective
creep at higher fields, $H>3$ T \cite{Blatter}. By contrast, the
activation energy for $H||ab$ exhibits a rather weak power law
decrease $E_m\propto H^{-0.17}$ in the entire field interval where
$\rho_{xx}(T,H)$ was measured.

The field and temperature dependencies of $E_a(T,H)$ can be
described by the following scaling function
    \begin{equation}
    E_a=\frac{E_0(1-T/T_c)^\alpha}{[1+H/H_0(T)]^\beta}\left[1-\frac{H}{H_{c2}(T)}\right]^\delta ,
    \label{Ea}
    \end{equation}
where $H_0$ quantifies the field above which vortex dynamics is
determined by collective effects of vortex interaction
\cite{Blatter}. From the data presented above, we obtain $H_0\simeq
3$ T, $\alpha\approx 1$, and $\beta\approx 1.1$ for ${\bf H}\| c$
and $\beta\approx 0.17$ for ${\bf H}\| ab$.

\section{Angular dependencies}

\begin{figure}[t]
\includegraphics[width=7.5cm]{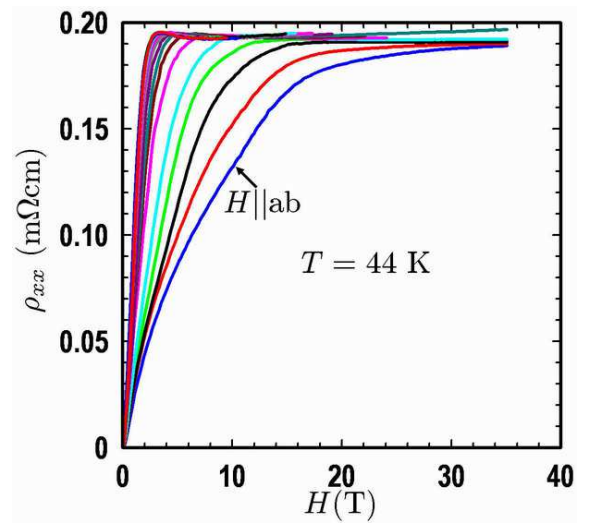}

\centering

\caption{The resistivities $\rho_{xx}(T,H)$ measured at 44~K for
different angles $\theta$ between $\bf{H}$ and the ab
plane.}\label{fig:angular}
\end{figure}

\begin{figure}[t]
\includegraphics[width=7.5cm]{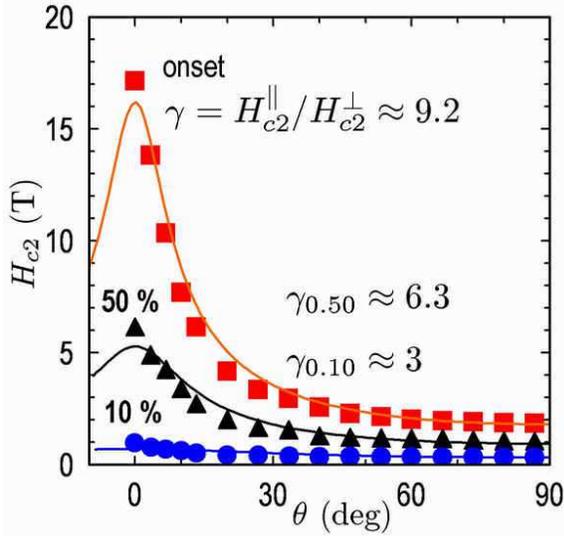}
\centering

\caption{Angular dependence of $H_{c2}(\theta)$ defined at
$\rho_{xx}=0.9\;0.5\;0.1\rho_n$ at 44~K. The solid lines show the
fits with Eq. (\ref{ang}). } \label{fig:gl}
\end{figure}

The TAFF resistivity $\rho_{xx}(T,H,\theta)$ was measured at
different angles $\theta$ between $\bf{H}$ and the ab plane, as
illustrated by a representative example shown in
Fig.~\ref{fig:angular}. The angular dependencies of $H_{c2}(\theta)$
extracted from the $\rho_{xx}(T,H,\theta)$ at 90~\%, 50~\%, and
10~\% of $\rho_{n}(T,H)$ are shown in Fig.~\ref{fig:gl}.

It  is clearly seen that the anisotropy  as $\gamma=
H_{c2}^{\|}/H_{c2}^{\perp}$ decreases when  calculated at lower
resistive transitions.  At the same time  $H_{c2}^{90\%}(\theta)$
shown in Fig.~\ref{fig:gl} notably departures from  the standard
Ginzburg-Landau formula:
    \begin{equation}
    H_{c2}(\theta)=\frac{H_{c2}^\perp}{\sqrt{\sin^2\theta+\gamma^{-2}\cos^2\theta}}
    \label{ang}
    \end{equation}
where $\gamma=(m_c/m_a)^{1/2}$ is defined by the ratio of
effective masses along the ab plane and the c-axis. The fit to
Eq.~(\ref{ang}) yields $\gamma\approx8.9$ in this case. Data taken
at 50~\% and 10~\% departure from Eq.~(\ref{ang}) even stronger,
which may result from increasing contribution of TAFF transport at
lower resistivities.

\begin{figure}[t]
\includegraphics[width=7.5cm]{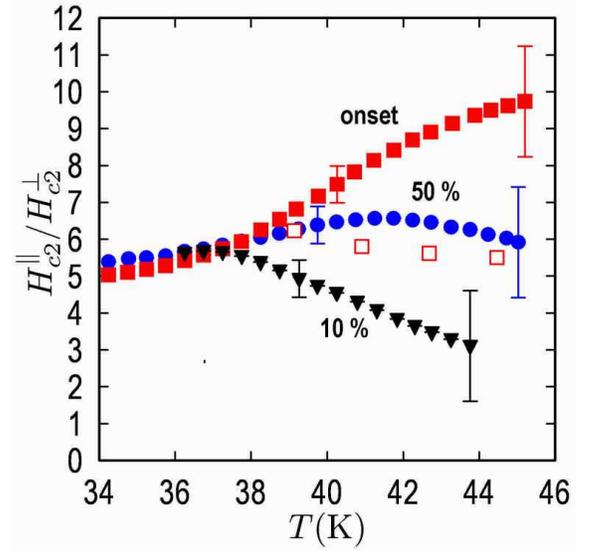}
\centering \caption{Temperature dependencies of  $\gamma_{10}$,
$\gamma_{50}$ and $\gamma_{90}$. The open squares show the data
from Ref.~35 measured on Nd(O$_{0.82}$F$_{0.18}$)FeAs single
crystal with higher $T_c=51.5$~K. These data are plotted at
reduced temperatures.}\label{fig:gamma}
\end{figure}

The temperature dependencies of $\gamma(T)$ obtained from the
$H_{c2}$ data shown in Figs.~\ref{fig:Bc2ab} and \ref{fig:Bc2c}
for $H_{10}$, $H_{50}$ and $H_{c2}$ are shown in Fig. 15. Contrary
to the standard one-band behavior, $\gamma$ turns out to be
temperature dependent, the values of $\gamma_{10}(T)$,
$\gamma_{50}(T)$ and $\gamma(T)$ being rather different,
especially just below $T_c$. While the parameter of $H_{c2}$
anisotropy, $\gamma(T)=H_{c2}^{||}/H_{c2}^{\perp}$ decreases from
$\approx 9$ at $T_c$ to $\approx 5$ at 34~K, the field $H_{10}$
becomes more anisotropic upon decreasing $T$. Below
$T\approx38$~K, however, these values of $\gamma(T)$ converge.

\section{$H_{c2}$ in two-band superconductors}

To understand how different pairing scenarios could manifest
themselves in the observed temperature dependencies of $H_{c2}(T)$,
we use the two-band BCS theory in which $T_c$ for interband
scattering is given by \cite{tg1,tg2}
    \begin{equation}
    T_{c0}=\Omega\exp[-(\lambda_+-s\lambda_0)/2w],
    \label{suhl}
    \end{equation}
where $\lambda_{\pm}=\lambda_{11}\pm\lambda_{22}$,
$w=\lambda_{11}\lambda_{22}-\lambda_{12}\lambda_{21}$,
$s=\mbox{sign}(w)$, and
$\lambda_0=(\lambda_-^2+4\lambda_{12}\lambda_{21})^{1/2}$,
$\lambda_{12}$ and $\lambda_{22}$ are pairing constants in bands 1
and 2, and $\lambda_{12}$ and $\lambda_{21}$ quantify interband
coupling. For s-wave pairing, intraband impurity scattering does not
affect $T_{c0}$, while weak interband scattering reduces $T_c$
according to \cite{imp1,imp2,ag}
    \begin{equation}
    T_c=T_{c0}-
    \frac{\pi}{8\lambda_0}[\lambda_0\gamma_++\lambda_-\gamma_--2\lambda_{21}\gamma_{12}-2\lambda_{12}\gamma_{21}]
    \label{tcsmall}
    \end{equation}
where $\gamma_\pm=\gamma_{12}\pm\gamma_{21}$, and $\gamma_{12}$ and
$\gamma_{21}$ are scattering rates between bands 1 and 2,
$N_1\gamma_{12}=N_2\gamma_{21}$ and
$N_1\lambda_{12}=N_2\lambda_{21}$, where $N_1$ and $N_2$ are partial
densities of states in bands 1 and 2, respectively. In the following
we discuss two scenarios: 1. Conventional s-wave pairing dominated
by strong intraband coupling, $w>0$. 2. Strong interband pairing
$w<0$ which can result in the $\pi$ shift between the order
parameters on two bands for interband repulsion $\lambda_{12}<0$
($s^\pm$ pairing \cite{mazin}).

As follows from Eq. (\ref{tcsmall}), interband repulsion with
$\lambda_{12} <0$ and $\lambda_{21}<0$ enhances $T_c$ depression due
to interband impurity scattering between different pieces of the
Fermi surface with opposite signs of the order parameter. This
suppression may be reduced by strong coupling effects \cite{imp3}.
By contrast, for strong interband attraction,
$4\lambda_{12}\lambda_{21}\gg \lambda_-^2$, Eq. (\ref{tcsmall})
yields
    \begin{equation}
    T_c=T_{c0}-\frac{\pi\gamma_{12}}{8}\left(1-\sqrt{\frac{N_1}{N_2}}\right)^2
    \label{tclim}
    \end{equation}
Here $T_c$ suppression is much weaker than for the $s^\pm$ pairing
because the factor $(1-\sqrt{N_1/N_2})^2$ reduces the effect of
interband scattering, making $T_c$ independent of $\gamma_{12}$ in
the special case of $N_1=N_2$. For instance, for $N_1=1.2N_2$, Eq.
(\ref{tclim}) gives a rather weak suppression of $T_c\simeq
T_{c0}(1-0.02g)$ even in the dirty limit, $g=\gamma_{12}/2\pi
T_{c0}>1$.

In the dirty limit, the equation for $H_{c2}$, which takes into
account both orbital and Zeeman pairbreaking for negligible
interband scattering, can be written in the following parametric
form \cite{ag}:
    \begin{eqnarray}
    \ln t=-[U_1(h)+U_2(h)+\lambda_0/w]/2+
    \\ \nonumber
    s[(U_1(h)-U_2(
    h)-\lambda_-/w)^2/4+\lambda_{12}\lambda_{21}/w^2]^{1/2},
    \label{hc2} \\
    H_{c2}=2\phi_0k_BT_cth/\hbar D_0,\qquad\qquad
    \label{hh2}
    \end{eqnarray}
where $t=T/T_{c0}$, and the functions $U_{1,2}$ are defined by
    \begin{equation}
    U_{1,2}(h)= \mbox{Re}\psi[1/2+(i+D_{1,2}/D_0)h] - \psi(1/2),
    \label{up}
    \end{equation}
$\psi(x)$ is the di-gamma function, $D_1$ and $D_2$ are
diffusivities in band 1 and 2, $D_0=\hbar/2m$, and the parameter
$h$ runs from $0$ to $\infty$ as $T$ varies from $T_c$ to 0. For
equal diffusivities, $\eta=D_2/D_1=1$, and negligible Zeeman
pairbreaking, $D_0\ll D_{1,2}$,  Eq. (\ref{hc2}) simplifies to the
one-band de-Gennes-Maki equation, $\ln t+U(h)=0$. If the magnetic
field H is inclined by the angle $\theta$ with respect to the ab
planes, Eqs. (\ref{hc2})-(\ref{up}) also describe the angular
dependence of $H_{c2}(\theta)$ if the intraband diffusivities are
replaced by their angular-dependent values:
    \begin{equation}
    D_m(\theta)=[D_m^{(a)2}\sin^2\theta+D_m^{(a)}D_m^{(c)}\cos^2\theta]^{1/2}
    \label{angl}
    \end{equation}
where $m=1,2$, and $D_m^{(a)}$ and $D_m^{(c)}$ are the in-plane and
the c-axis principal values of $D_m^{\alpha\beta}$.

$H_{c2}(T)$ curves shown in Fig.~\ref{fig:Bc2ab} and
\ref{fig:Bc2c} exhibit pronounced upward curvature for the field
along the c-axis, and the downward curvature for the field along
the ab plane, similar to the behavior of $H_{c2}(T)$ first
observed on polycrystalline LaO$_{0.89}$F$_{0.11}$FeAs
\cite{Hunte}. Such temperature dependence of $H_{c2}(T)$ has been
often observed on MgB$_2$, suggesting that the two-band theory in
which two bands have different diffusivities can also be used to
describe $H_{c2}(T)$ in oxypnictides (the superconducting gaps on
the disconnected pieces of the Fermi surface can be either
different or the same). If interband scattering is negligible,
neither $T_c$ nor $H_{c2}$ depend on the sign of $\lambda_{12}$,
so we consider two different cases: 1. dominant intraband
coupling, $w=\lambda_{11}\lambda_{22}-\lambda_{12}\lambda_{21}>0$
and 2. dominant interband coupling,
$w=\lambda_{11}\lambda_{22}-\lambda_{12}\lambda_{21}<0$.

\begin{figure*}
\includegraphics[width=17.0cm]{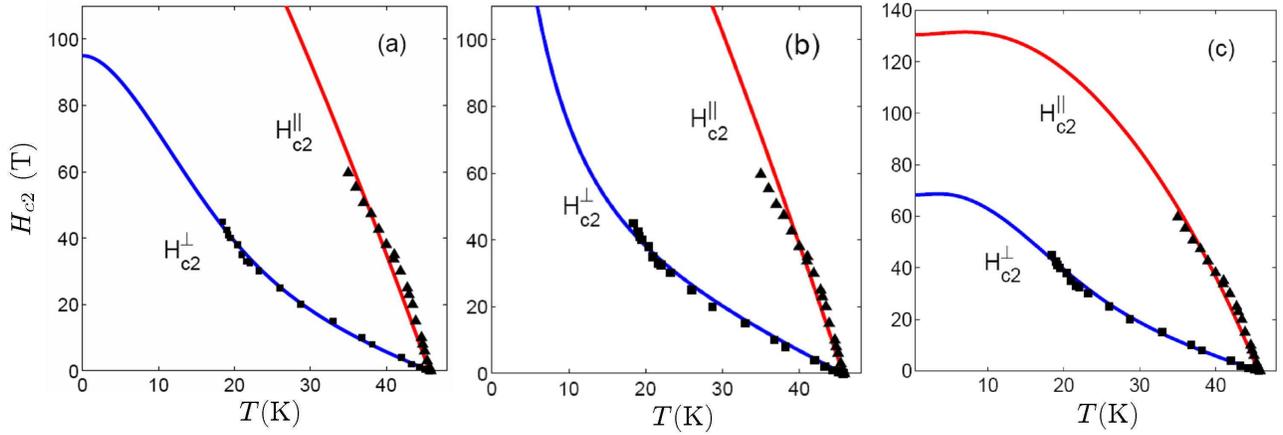}
\centering
\centering
%\includegraphics[width=8.5cm]{ndc.eps}
%\centering

\caption{Fits of Eqs. (\ref{hc2})-(\ref{up}) to the experimental
data shown in Figs.~\ref{fig:Bc2ab} and \ref{fig:Bc2c} for different
pairing scenarios: $w>0$, $\lambda_{11}=\lambda_{22}=0.5$,
$\lambda_{12}=\lambda_{21}=0.25$,
$\eta_\perp=D_2^{(a)}/D_1^{(b)}=0.08$,
$\eta_{||}=[D_2^{(a)}D_2^{(c)}/D_1^{(a)}D_1^{(c)}]^{1/2}=1$, and
negligible paramagnetic effects (a); $w<0$,
$\lambda_{11}=\lambda_{22}=0.49$, $\lambda_{12}=\lambda_{21}=0.5$,
$\eta_\perp=D_2^{(a)}/D_1^{(b)}=0.007$,
$\eta_{||}=[D_2^{(a)}D_2^{(c)}/D_1^{(a)}D_1^{(c)}]^{1/2}=1$, and
negligible paramagnetic effects (b); $w>0$,
$\lambda_{11}=\lambda_{22}=0.5$, $\lambda_{12}=\lambda_{21}=0.25$,
$\eta_\perp=D_2^{(a)}/D_1^{(b)}=0.08$,
$\eta_{||}=[D_2^{(a)}D_2^{(c)}/D_1^{(a)}D_1^{(c)}]^{1/2}=1$,
$d_\perp=D_1^{(a)}/D_0=10$, and
$d_{||}=[D_2^{(a)}D_2^{(c)}/D_1^{(a)}D_1^{(c)}]^{1/2}/D_0=0.7$ (c).}
\label{fig:fit}
 \end{figure*}

Shown in Fig.~\ref{fig:fit}(a) is an example of the fit of
Eq.~(\ref{hc2})-(\ref{up}) to the data for  case 1 with
$\lambda_{11}=\lambda_{22}=0.5$,  and
$\lambda_{12}=\lambda_{21}=0.25$, in which case
$T_c=\Omega\exp[-1/(\lambda_{11}+\lambda_{12})]=50$K corresponds
to $\Omega\simeq 190$K. Fig.~\ref{fig:fit}(a) shows the fit
without taking  paramagnetic effects into account for
$D_2=0.08D_1$, which suggests that scattering in one band is
significantly stronger than in the other. We do not discuss here
microscopic mechanisms behind such band disparity; instead we
focus on the set of parameters which fit the data and the extent
to which these parameters are different for the case of $w>0$ and
$w<0$. Extrapolations of $H_{c2}^{\|}(T)$ to $T=0$ based on the
observed slope at $T_c$ suggests
$H_{c2}^\|(0)\sim\mathrm{\mathrm{\mathrm{}}} 220$T, well above the
weak coupling BCS paramagnetic limit $H_p[$T$]=1.84T_c[$K$]\simeq
90$T.

Fig.~\ref{fig:fit}(a) shows the fit in the case of strong
interband pairing
$\lambda_{12}\lambda_{21}>\lambda_{11}\lambda_{21}$ for
$\lambda_{11}=\lambda_{22}=0.49$, $\lambda_{12}=\lambda_{21}=0.5$
and $D_2^{a}=0.007D_1^{a}$. Reproducing the observed upward
curvature of $H_{c2}^\perp(T)$ for this case requires a much
greater disparity of the band diffusivities than for the previous
case. Reducing $\lambda_{11}$ and $\lambda_{22}$ shifts the region
of upward curvature to lower temperatures, which makes fitting the
data even more difficult. These features may put serious
constraints on the $s^\pm$ pairing scenario.

Given the very high extrapolated values of $H_{c2}^\|(0)$ in
Fig.~\ref{fig:fit}(a), the role of paramagnetic effects becomes very
important. Shown in Fig.~\ref{fig:fit}(c) is the fit for the same
parameters as in Fig.~\ref{fig:fit}(a) but with the account of the
Zeeman pairbreaking for $D_1^{(a)}=10D_0$ and
$[D_2^{(a)}D_2^{(c)}/D_1^{(a)}D_1^{(c)}]^{1/2}=0.7D_0$. In this case
the values of $H_{c2}(T)$ at low temperatures are significantly
reduced as compared to what may be expected from the fit in
Fig.~\ref{fig:fit}(a), which ignores the paramagnetic effects.

Several remarks should be made regarding the fits in
Fig.~\ref{fig:fit}(a-c). First, applying weak coupling
Eqs.~(\ref{hc2})-(\ref{up}) for a superconductors with $T_c=50$~K
obviously neglects the strong coupling renormalization effects of
the Eliashberg theory. However given the lack of microscopic theory
of superconductivity in oxypnictides, the use of the BCS-type
approach may have its own merits given that the shape of the
$H_{c2}(T)$ curve described by Eqs. (\ref{hc2})-(\ref{up}) depends
only on the ratio of the constants $\lambda_{mn}$ but it is
basically insensitive to the particular s-wave pairing mechanisms.
The usual way of treating strong coupling effects in Eqs.
(\ref{hc2})-(\ref{up}) is to express them in terms of experimentally
observed parameters \cite{renorm} such as $T_c$ and $D_1$ and $D_2$
extracted from normal state transport measurements and the observed
slopes $dH_{c2}/dT$. The second remark is that the actual
paramagnetic limit $H_p\simeq (1+\lambda)H_p^{BCS}$ is significantly
enhanced by strong coupling effects as compared to the BCS value
\cite{para}. Here $\lambda\sim 1$ is a characteristic bare coupling
constant for the relevant exchange boson, so $H_{c2}(T)$ values may
significantly exceed $H_p^{BCS}$ as has been observed in other
superconductors, for example PbMo$_6$S$_8$ \cite{chevr1,chevr2}.
Thus, Eqs. (\ref{hc2})-(\ref{up}) may capture the qualitative
behavior of $H_{c2}(T)$ as functions of materials parameters if they
are expressed in terms of the observed quantities, and the Zeeman
terms are renormalized by strong coupling effects.

\section{Discussion}

Our high-field data obtained on a Nd(O$_{1-x}$F$_x$)FeAs single
crystals show extremely high $H_{c2}$ values, consistent with
previous low-field measurements on single crystals
\cite{lowfield1,Wen_heat,Wen_Hall,Wen_Tc,Wen_anizo} and high-field
measurements on polycrystals \cite{Hunte,jjprb08}. Our values of
the temperature-dependent $H_{c2}$ slopes, $H_{c2}^{\prime\|}
\approx 8-10$~T/K and $H_{c2}^{\prime\perp} \approx 1.2-5$~T/K
measured from the resistivity onset of superconductivity are
considerably higher than the values of $H_{c2}$ extracted from
calorimetry measurements on Nd(O$_{1-x}$F$_x$)FeAs single crystals
\cite{Wen_heat}, which may be affected by strong paramagnetism of
Nd$^{3+}$ ions at high fields \cite{mag2}.  We should also mention
several issues, which may complicate accurate extraction of the
anisotropy in Nd(O$_{1-x}$F$_x$)FeAs single crystals grown by the
flux method, which usually results in intergrowths and stacking
faults parallel to the ab plane. Such planar defects may increase
the apparent anisotropy parameter $\gamma$ and give rise to the
local modulations of order parameter along the c-axis. The latter
may broaden the resistive transition at $T_c$  and the upward
curvature of $H_{c2}(T)$ near $T_c$, which could mask the two-band
effects.

Because of very high values of $H_{c2}^{\prime\|}$ and
$H_{c2}^{\prime\perp}$, our high field measurements up to 60~T are
still mostly limited to the temperatures not too far from $T_c$, particularly for
the field parallel to the ab plane. As a result, it is hard to
unambiguously evaluate the values $H_{c2}(0)$, as
Figs.~\ref{fig:fit}(a-c) illustrate. Yet we can evaluate the
in-plane and the c-axis GL coherence lengths, $\xi_a=(\phi_0/2\pi
H_{c2}^{\prime\perp} T_c)^{1/2}\simeq 2.3$~nm and $\xi_c =
\xi_a/\gamma = 0.26$~nm for $\gamma = 9$, $H_{c2\perp}^\prime
=1.2$~T/K and $T_c=50$~K. These values of $\xi_a$ and $\xi_c$ are
not too different from those for YBa$_2$Cu$_3$O$_{7-x}$. The fits
shown in Figs.~\ref{fig:fit}(a-c) also indicate the importance of
paramagnetic effects in $H_{c2}(T)$, particularly the fact that
extrapolations of $H_{c2}$ to low temperatures based on the measured
slope $H_{c2}^\prime$ and the WHH formula $H_{c2}\simeq
0.7T_cH_{c2}^\prime$ \cite{lowfield3} may grossly overestimate
$H_{c2}(0)$. At the same time, the importance of paramagnetic
effects in oxypnictides might open up possibilities of observing
first order phase transitions or perhaps the
Fulde-Ferrel-Larkin-Ovchinnikov inhomogeneous states at very high
magnetic fields.

The temperature dependence of $H_{c2}^\perp(T)$ shown in
Fig.~\ref{fig:Bc2c} is rather different from the conventional
one-band WHH behavior, which may be characteristic of many
oxypnictides. This confirms our earlier qualitative conclusion
inferred from indirect measurements of $H_{c2}^\perp(T)$ on
polycrystalline oxypnictides \cite{Hunte}. One possibility to
explain the significant upward curvature of $H_{c2}^\perp(T)$ is
based on the two-band model. This conclusion is also consistent
with the temperature dependent mass anisotropy parameter,
$\gamma(T)$, reminiscent of the behavior of MgB$_2$
\cite{ag,mgb2}. Yet many important points remains controversial.
First, to explain the observed upward curvature of
$H_{c2}^\perp(T)$, we had to assume a rather high difference in
the intraband diffusivities, $D_2\sim (0.1-0.01)D_1$ (depending on
the pairing scenario), considerably higher than the effective mass
difference for $\Gamma$ and $M$ electron pockets predicted by
ab-initio calculations \cite{theor1,theor2,theor3,theor5}. One has
therefore to assume that scattering on impurities or strong
magnetic excitations may produce such big differences in $D_1$ and
$D_2$.

Our data enable us to make further qualitative conclusions
regarding the s-wave two-band pairing scenarios with and without
the interband $\pi$ shift. The first one is less susceptible to
the pairbreaking effect of interband impurity scattering, as
follows from Eqs. (\ref{tcsmall}) and (\ref{tclim}). Indeed, the
fact that our Nd(O,F)FeAs single crystal has the same $T_c$ as
 sintered multiphase polycrystals, seems to suggest that the
impurity scattering does not suppress $T_c$ much. The latter may
also indicate that oxypnictides are effectively in the clean limit
just because of their short coherence lengths, $\xi_a <\ell$, where
$\ell$ is the mean free path. Yet $s^\pm$ pairing is certainly
more susceptible to the pairbreaking interband impurity
scattering. Another indication that the s-wave pairing without the
interband $\pi$ shift appears to fit our data better comes from
the fact that, in order to explain the observed upward curvature
of $H_{c2}^\perp(T)$ in Fig.~\ref{fig:fit}(b) in the $s^\pm$
model, we have to assume the ratio $D_1/D_2$ to be about an order
of magnitude smaller that $D_2/D_1\sim 0.1$ required for the more
traditional model of strong intraband pairing and weaker interband
pairing. These constraints may reduce the number of possible
scenarios of superconductivity in the oxypnictides.

As far as vortex dynamics in Nb(F,O)FeAs is concerned, our data
presented in Figs.~\ref{fig:rho3} and  \ref{fig:rho4} show that
high magnetic fields parallel to the c-axis mostly shift the
resistive transition without significant broadening of the
$\rho_{xx}(T)$ curves, while  $H$ parallel to the ab plane changes
the shape of $\rho_{xx}(T)$, but certainly not as strong as for
high-$T_c$ cuprates \cite{Blatter}. This suggests that despite the
rather high values of $\gamma$, thermal fluctuation of vortices in
Nb(F,O)FeAs single crystals are weaker than in most anisotropic
layered cuprates, like Bi-2212. This conclusion is consistent with
the data of other groups \cite{creepSm} and with the relatively
high activation energy $E_0\sim 3\times 10^3$~K extracted from our
transport measurements. The effect of thermal fluctuations is also
quantified by the Ginzburg parameter, $Gi = (2\pi
k_BT_c\mu_0\Lambda_0^2/\phi_0^2\xi_c)^2/2$, where $\Lambda_0$ is
the in-plane London penetration depth. Taking $\Lambda_0=200$ nm,
$\xi_c=0.26$~nm and $T_c=49$~K, we obtain $Gi\simeq 10^{-2}$, of
the order of the typical Ginzburg number for YBCO. For two-band
superconductors, the above estimates of $Gi$ remain qualitatively
the same if $\gamma$ and $\xi$ are taken for the band with the
minimum effective mass or maximum electron mobility \cite{ag}. By
contrast, $Gi$ in La(O,F)FeAs is close to $Gi$ in MgB$_2$, about
30 times smaller than $Gi$ in Nd(O,F)FeAs \cite{Hunte,jjprb08}.

In conclusion, our high-field magneto-transport measurements on
single crystal Nd(O$_{0.70}$F$_{0.30}$)FeAs have revealed very
high upper critical fields and their anomalous temperature
dependencies. We also observe a pronounced thermally-activated
flux flow resistivity which may indicate rich vortex dynamics in
single layer oxypnictides.

The work at NHMFL was supported by the NSF Cooperative Agreement No.
DMR-0084173, by the State of Florida,  by the DOE, by the NHMFL IHRP
program (FH), and by AFOSR grant FA9550-06-1-0474 (AG and DCL).

%\bibliography{oxy3}

\end{document}